\documentclass{PoS}
\usepackage{aas_macros}

\title{BurstCube:  A CubeSat for Gravitational Wave Counterparts}

\ShortTitle{BurstCube}

\author{\speaker{Judith Racusin}\\
        NASA/GSFC\\
        E-mail: \email{judith.racusin@nasa.gov}}

\author{Jeremy S. Perkins\\
        NASA/GSFC\\
        E-mail: \email{jeremy.s.perkins@nasa.gov}}
        
\author{Michael S. Briggs$^a$, Georgia de Nolfo$^b$, John Krizmanic$^c$, Regina Caputo$^c$, Julie E. McEnery$^b$, Peter Shawhan$^d$, David Morris$^e$, Valerie Connaughton$^f$, Dan Kocevski$^g$, Colleen Wilson-Hodge$^g$, Michelle Hui$^g$, Lee Mitchell$^h$, \& Sheila McBreen$^i$\\
\llap{$^a$}University of Alabama, Huntsville\\
\llap{$^b$}NASA/GSFC\\
\llap{$^c$}NASA/GSFC/CRESST\\
\llap{$^d$}University of Maryland, College Park\\
\llap{$^e$}University of the Virgin Islands\\
\llap{$^f$}USRA\\
\llap{$^g$}NASA/MSFC\\
\llap{$^h$}Naval Research Laboratory\\
\llap{$^i$}University College Dublin\\
}

\abstract{We present BurstCube, a novel CubeSat that will detect and localize Gamma-ray Bursts (GRBs). 

BurstCube will detect long GRBs, attributed to the collapse of massive stars, short GRBs (sGRBs), resulting from binary neutron star mergers, as well as other gamma-ray transients in the energy range 10-1000 keV. sGRBs are of particular interest because they are predicted to be the counterparts of gravitational wave (GW) sources soon to be detectable by LIGO/Virgo. BurstCube contains 4 CsI scintillators coupled with arrays of compact low-power Silicon photomultipliers (SiPMs) on a 6U Dellingr bus, a flagship modular platform that is easily modifiable for a variety of 6U CubeSat architectures.  BurstCube will complement existing facilities such as Swift and Fermi in the short term, and provide a means for GRB detection, localization, and characterization in the interim time before the next generation future gamma-ray mission flies, as well as space-qualify SiPMs and test technologies for future use on larger gamma-ray missions. The ultimate configuration of BurstCube is to have a set of $\sim10$ BurstCubes to provide all-sky coverage to GRBs for substantially lower cost than a full-scale mission.}

\FullConference{35th International Cosmic Ray Conference --- ICRC2017\\
		10--20 July, 2017\\
		Bexco, Busan, Korea}

\begin{document}

\section{Introduction}
The first direct detections of Gravitational Waves (GWs) by the Laser Interferometer GW Observatory (LIGO) of GW150914, GW151226 and GW170104, the mergers of stellar-mass black hole (BH) binaries \cite{gw150914,gw151226,PhysRevLett.118.221101}, has brought GW astronomy into a new era of discovery.  The search for electromagnetic (EM) counterparts to GW sources is now more important than ever before, in order to provide astrophysical context, and validate low-significance signals.  Prior to the discovery of GW150914, most of the theoretical predictions for EM counterparts to GW sources were for compact object mergers involving a neutron star (NS; i.e. NS-NS and NS-BH).  However, a candidate $\gamma$-ray counterpart that is both temporally and spatially coincident with GW150914 was detected by the \textit{Fermi} Gamma-ray Burst Monitor (GBM) \cite{2016arXiv160203920C}, and is consistent with being a low-fluence short $\gamma$-ray burst (sGRB), providing new hope for detecting EM counterparts to BH-BH mergers \cite{2016arXiv160204542Z,2016arXiv160204460L,2016arXiv160204735L,2016arXiv160205050Y,2016arXiv160205140P}.

LIGO and Virgo are commissioning major upgrades to reach new design sensitivities by 2019, adding the detectability of GWs from the inspiral of systems including a NS, generally believed to be the progenitors of sGRBs \cite{2013arXiv1304.0670L}.  The simultaneous discovery of GW and EM signatures requires dedicated and coordinated observations by large communities of both ground and space-based observatories.  Existing sensitive $\gamma$-ray burst (GRB) observatories
cover only $\sim70\%$ of the sky at any one time, and any increase in sky coverage by additional facilities increases both the likelihood of coincident detection, and the number of sGRBs that can be correlated with 
GW signals.

The recent detection of GW signals from binary BH mergers is the most important discovery of this decade so far.  It may be surpassed in the coming years by further GW discoveries, especially NS binary mergers which may reveal the mechanisms of most sGRBs and nuclear astrophysics. The GBM candidate coincident sGRB only adds to the need for EM searches for GW counterparts. The Advanced LIGO and Advanced Virgo GW observatories (hereafter LIGO/Virgo) are scheduled to reach
design sensitivity in 2019, with LIGO-India to be added in 2022 \cite{2013arXiv1304.0670L}.  The current fleet of sensitive spaceborne GRB instruments with localization capability has incomplete sky coverage
and uncertain operation beyond 2019.  {\bf Coincident detection of sGRBs and GW sources will:}
\begin{itemize}
\item provide the ``smoking gun'' evidence for the progenitors of
  sGRBs for the first time;
\item increase the confidence in low-significance LIGO/Virgo GW detections;
\item provide small temporal and positional windows for
targeted searches of GW data that will enable detecting
weaker GW signals than can be detected in blind searches
lacking priors, increasing the search volume;
\item provide the astrophysical context for the GW signal via
  population statistics on jet beaming angles and $\gamma$-ray
  energetics as inputs into stellar population synthesis models;
\item provide localizations that will assist wide-field follow-up observers in afterglow
detection and redshift measurement which will lead to insight
into cosmological parameter estimation, constraints on the NS equation of state \cite{2013PhRvL.111g1101D}, and an inventory of
r-process elements in the Universe constrained by the faint sGRB kilonova
signature \cite{2010MNRAS.406.2650M}.
\end{itemize}

\section{BurstCube}
Over the past decade there has been a renaissance in the development of small-sat and CubeSat technologies.  The result is that it is achievable to build a CubeSat to detect the bright signals from GRBs.  These proceedings detail a 6U CubeSat concept called BurstCube based on the Dellingr platform developed at GSFC but the overall concept can be generalized and use either a custom or commercial bus and be larger or smaller than BurstCube.  The main limiting factors in developing a mission like BurstCube are mass, power, and volume but the advent of low-power, low-volume SiPMs enables a workable detector readout design.  We describe the BurstCube CubeSat, which provides rapidly available high-resolution temporal, spectral, and localization data.  Even rough localizations will be helpful to verify weak GW signals, as the temporal coincidence is the most important measurement alerting the community to a potential on-axis event. 
BurstCube increases the sky coverage beyond existing
facilities, and is optimized for detecting sGRBs, all for a small
fraction of the cost of a larger mission.

\begin{figure}[t!]
\centering
\begin{minipage}{0.8\textwidth}
\includegraphics[width=\textwidth,trim=20 150 20 85,clip=true]{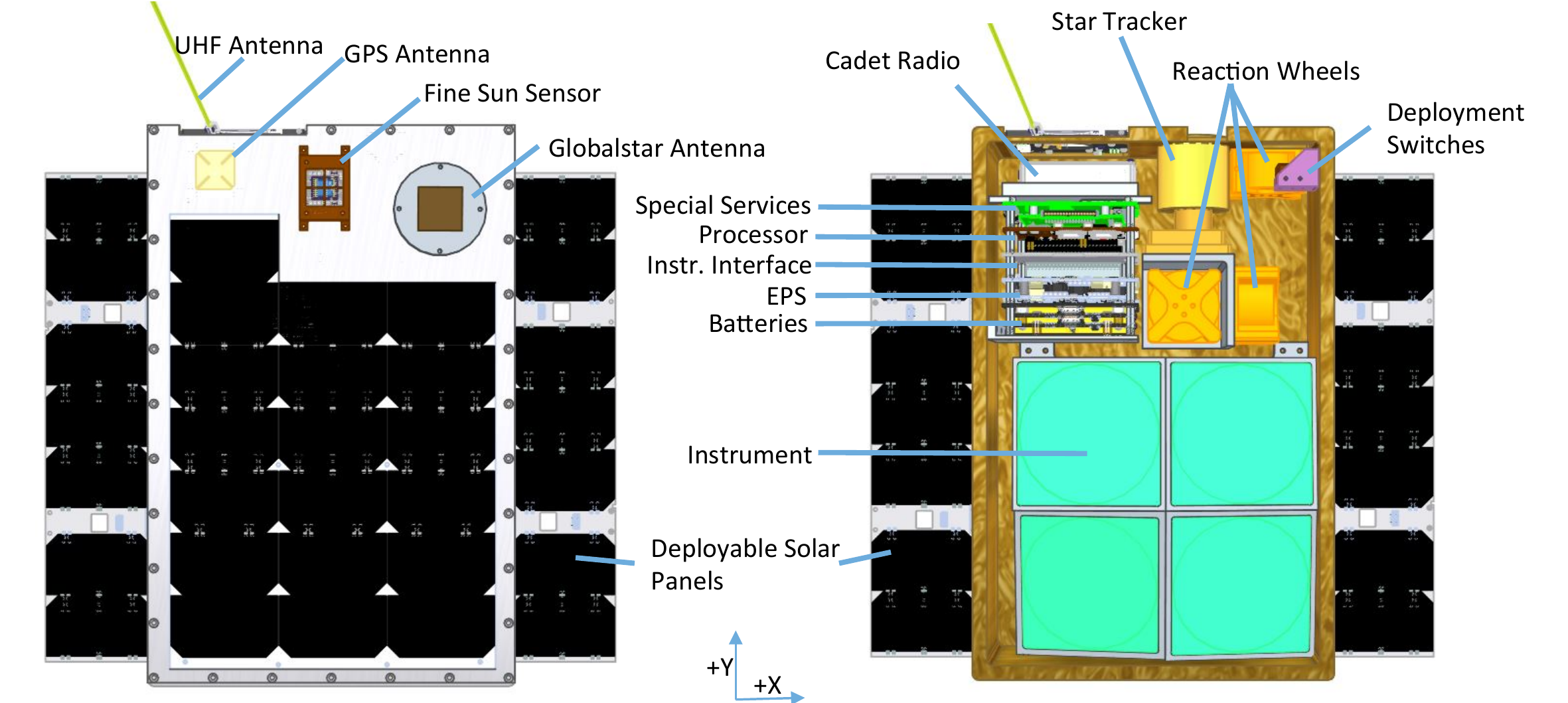}
\end{minipage}\hfill
\begin{minipage}{0.2\textwidth}
\caption{\small Internal view of BurstCube instrument and spacecraft components ({\it right}).  BurstCube body-mounted and deployable solar panels ({\it left}).
\label{fig:spacecraft}}
\end{minipage}
\end{figure}

BurstCube is a 6U CubeSat divided into a 4U instrument package, and 2U of spacecraft subsystems ({\bf Fig. \ref{fig:spacecraft}}).  The instrument concept is similar to
{\it Fermi}-GBM \cite{2009ApJ...702..791M}, except BurstCube uses CsI as a primary detection medium covering the energy range from 10 keV - 1 MeV with high efficiency and adequate energy resolution. These crystals are inexpensive and have a long track record of use in $\gamma$-ray astronomy. A design philosophy is to rely on proven well-tested technology.

Each of the four detectors are 9.4$\times$9.4$\times$1.27 cm, with the maximum dimension dictated by the size of the 6U CubeSat. The detectors are composed of CsI crystals viewed by an array of low-power and low-voltage (20-70 V) Silicon Photomultipliers (SiPMs).  Compared the more conventional photomultiplier tubes, SiPMs significantly reduce mass, volume, power, and cost.  The combination of scintillation crystals and new readout devices makes it possible to consider a compact, low-power instrument that is readily deployable on a CubeSat platform.  The format of the 6U spacecraft drives the instrument mass, power, and volume budget. However, even in this compact design the detector presented here is competitive with the state of the art system (\textit{Fermi}-GBM).  The BurstCube team is exploring a modification to the baseline design by further segmenting the detector system and triggering over summed detectors to better localize bright GRBs.

\begin{figure}[t!]
\vspace{-2mm}
\centering
\begin{minipage}{0.52\textwidth}
\centering
\includegraphics[width=0.8\textwidth,trim=20 0 50 35,clip=true]{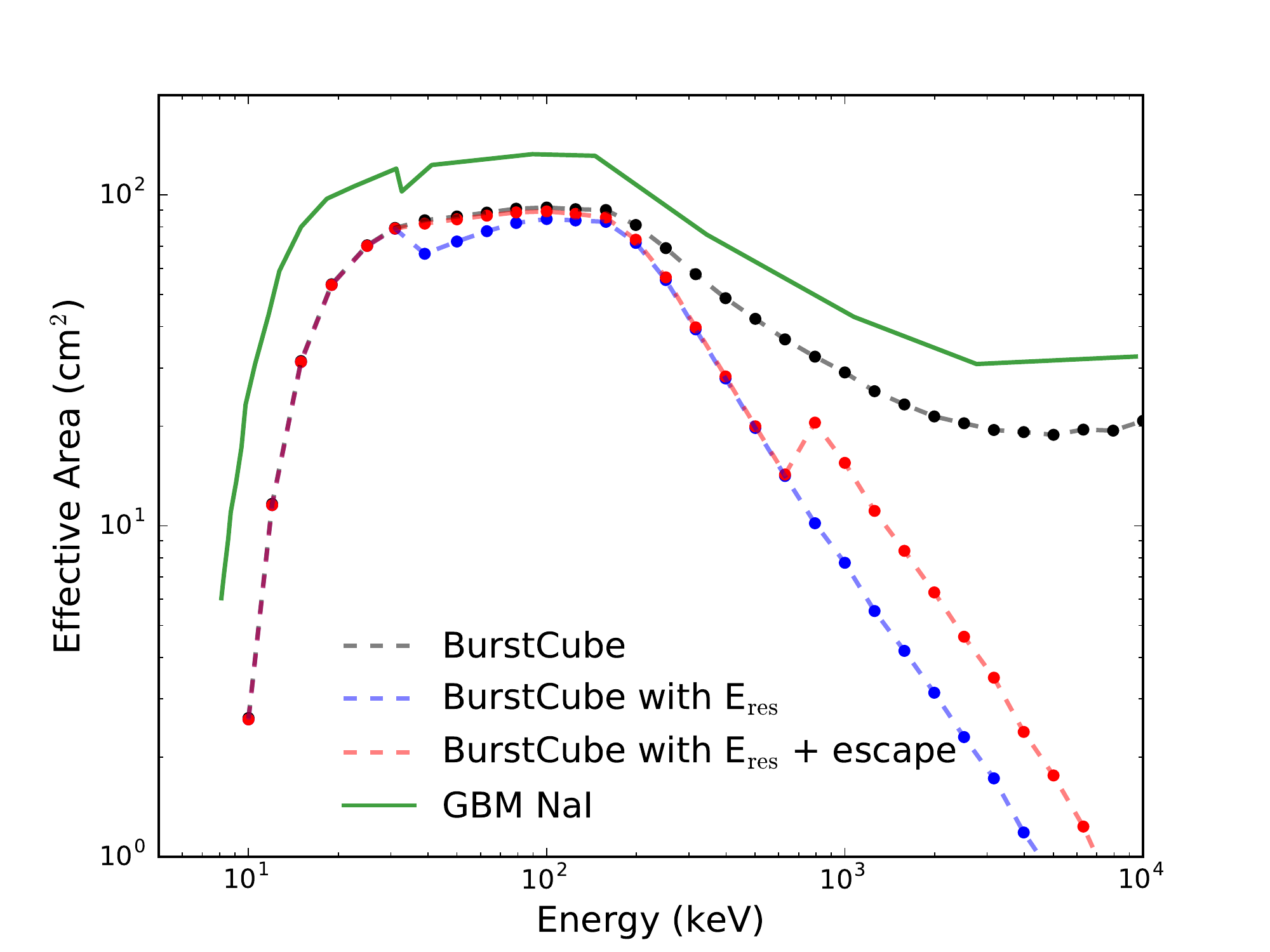}
\caption{\small Despite the constraints of a CubeSat, BurstCube achieves an effective area
of 70\% of GBM at 100 keV and $15^\circ$ incidence 
The effective area as a function of energy, and the corresponding
curve for the larger GBM NaI detectors are shown for reference.\label{fig:effarea_energy}}
\end{minipage}\hfill
\begin{minipage}{0.44\textwidth}
\vspace{-2mm}
\includegraphics[width=1.05\textwidth,trim=20 0 0 25,clip=true]{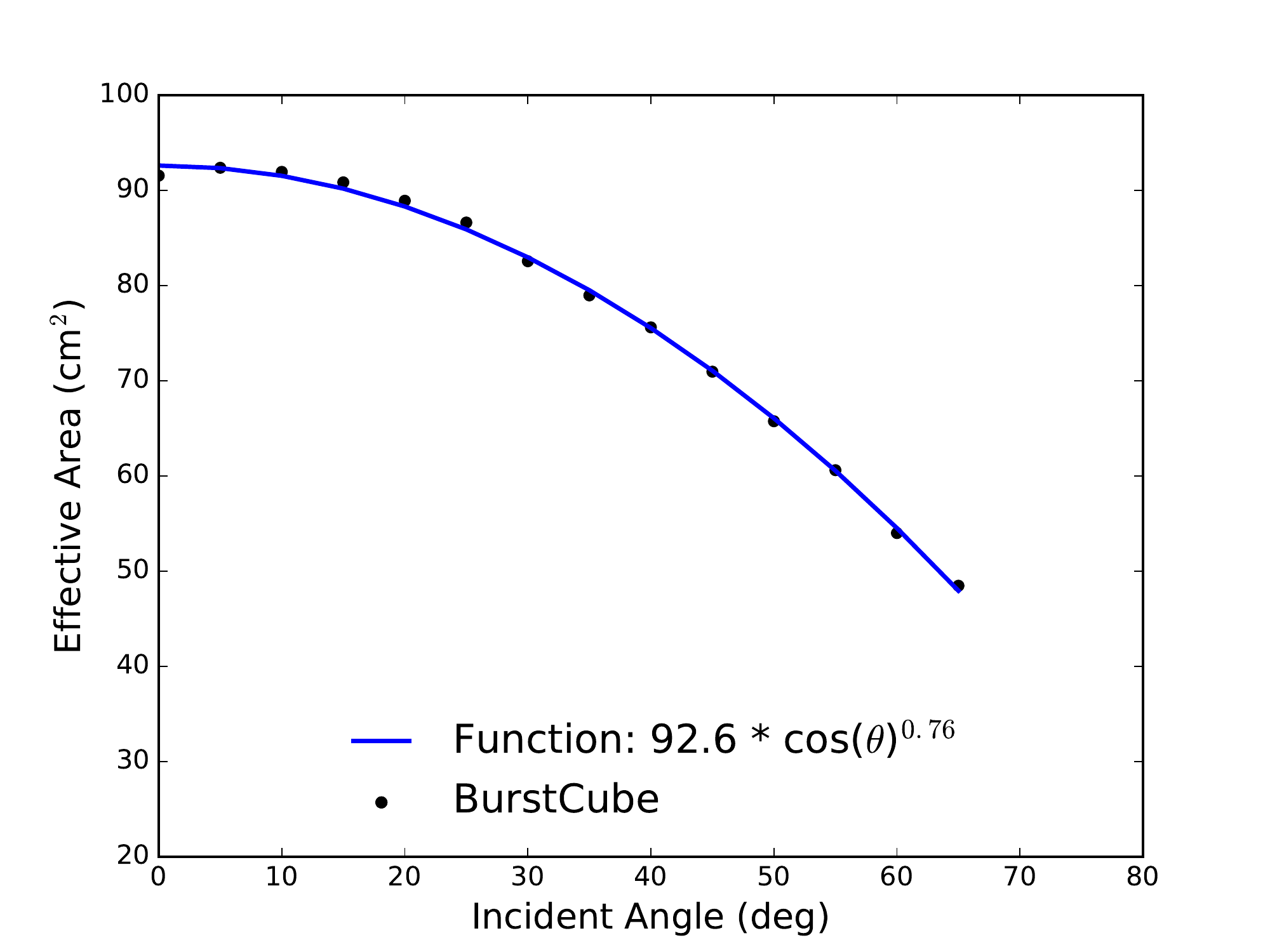}
\caption{\small The square BurstCube detectors make maximum use of
  space and still achieve an effective area cosine dependence on
  incidence angle expected for a thin detector which is needed for GRB localizations.
\label{fig:effarea_angle}}
\end{minipage}
\vspace{-5mm}
\end{figure}

To estimate the BurstCube effective area and resulting sensitivity, we simulated a single CsI detector using MEGAlib \cite{Megalib} (which provides an interface to GEANT4). The simulated BurstCube effective area is competitive with GBM ({\bf
  Fig. \ref{fig:effarea_energy}}), despite the smaller detectors. 

\begin{figure}[b!]
\vspace{-4mm}
\centering
\begin{minipage}{0.48\textwidth}
\includegraphics[width=\textwidth,trim=110 15 110 30,clip=true]{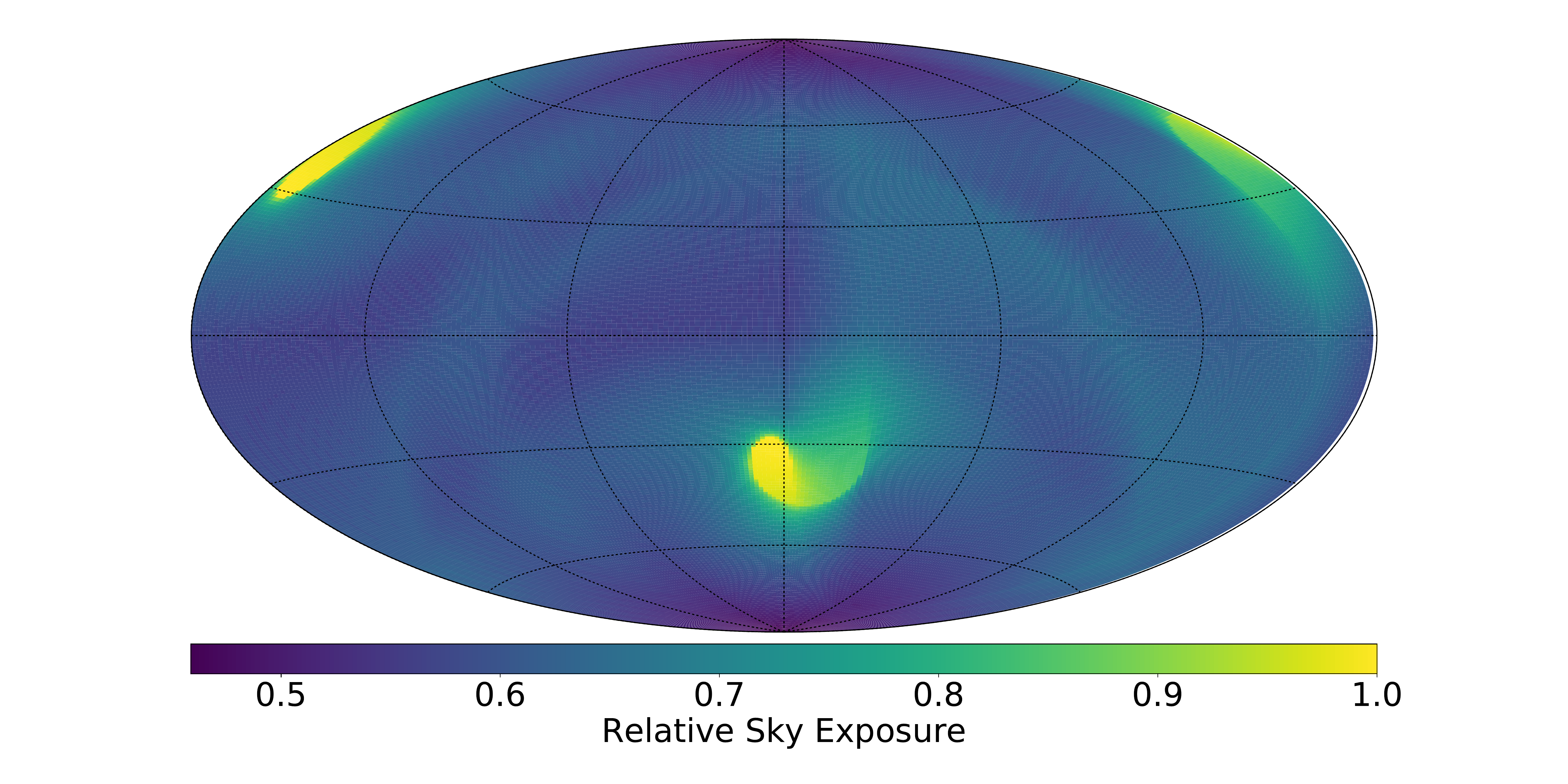}
\caption{\small The integrated relative exposure of BurstCube
over a single orbit projected onto the sky in Mollweide celestial coordinates.  An exposure of 1 is the maximum on-axis exposure for a
single CsI detector.\label{fig:skymaps}}
\end{minipage}\hfill
\begin{minipage}{0.48\textwidth}
  \includegraphics[width=\textwidth,trim=0 15 0 50,clip=true]{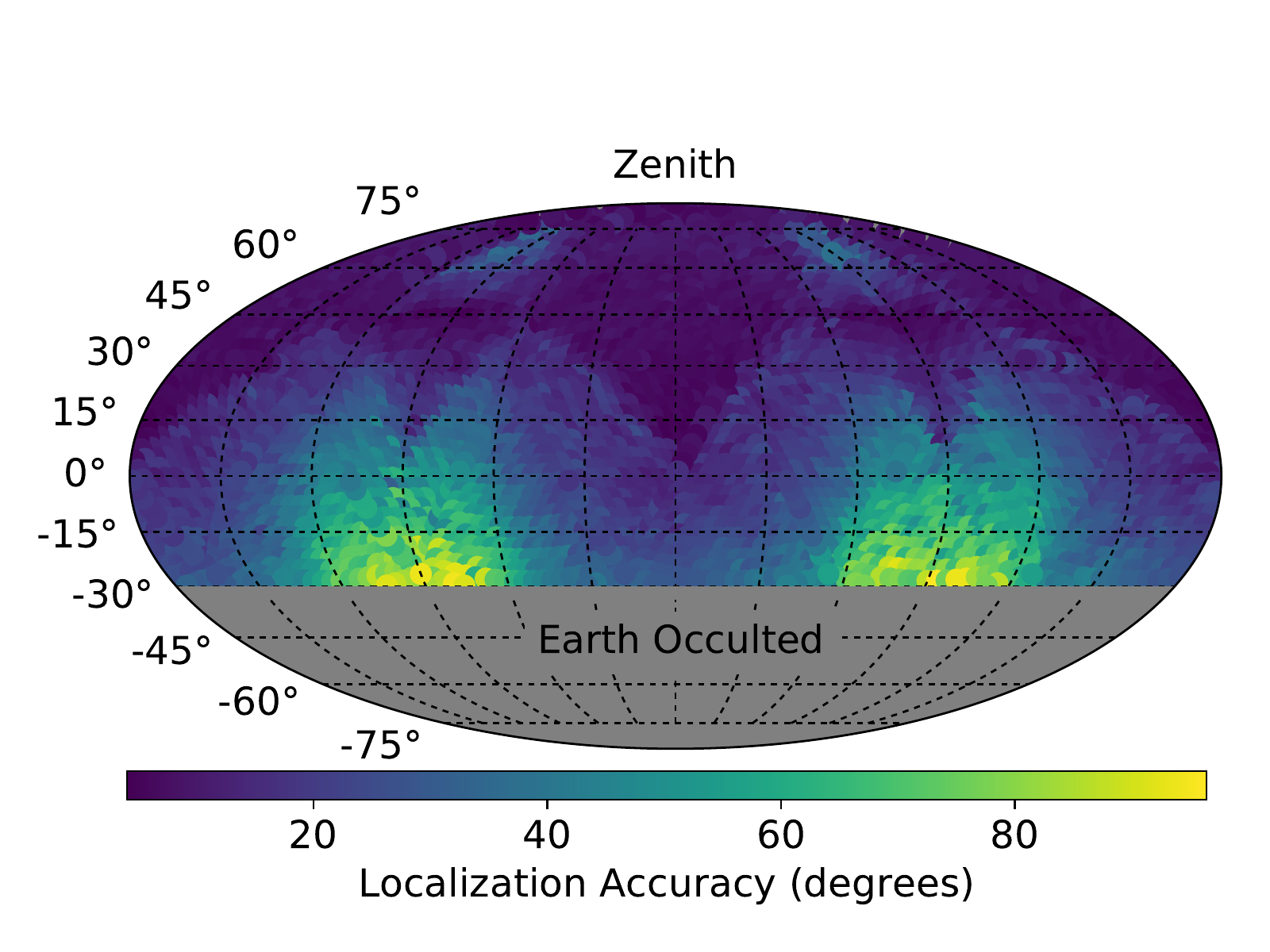}
  \caption{\small BurstCube localizes sGRBs to an accuracy of $\sim$7$^\circ$ radius when viewed by 3 or more detectors. The localization errors (in degrees radius) are shown across the entire BurstCube FoV in Az/El coordinates.  \label{fig:skymaps_instant}}
\end{minipage}
\vspace{-3mm}
\end{figure}

BurstCube will localize GRBs in a similar manner
to BATSE or GBM \cite{batse}. The approximate cosine dependence ($\sim$cos$(\theta)^{0.76}$) of the
effective area with incidence angle ({\bf Fig.
  \ref{fig:effarea_angle}}) is leveraged to localize GRBs by measuring
the relative brightness of the burst in each detector. The detector count rates will be matched to a previously computed table of
relative detector responses as a function of spacecraft coordinates.
Depending on the region of the sky in which the burst occurs,
the localization accuracy can be as good as a few degrees or as high
as 90 degrees.  The localization accuracy for each position on the sky
in Azimuth and Elevation coordinates ({\bf Figs.
  \ref{fig:skymaps} and \ref{fig:skymaps_instant}}), shows that over the $\sim1/3$
of the sky seen with 3 or 4 detectors, localizations have an typical accuracy of $<7^\circ$ radius.

{\bf BurstCube will detect $\gtrsim$ 24 sGRBs per year, in addition to $>$100 long GRBs and other transients.}  GRB rate estimates are made
by scaling the GBM GRB detections \cite{2014ApJS..211...13V} to
BurstCube based upon comparison between simulations of both the
BurstCube and GBM NaI detector systems.  We estimate the BurstCube
limiting flux for the various trigger timescales in the
50-300 keV band, and compare those to the similar GBM distributions to
determine which GBM sGRBs would be detectable by BurstCube.  These
calculations use only a few of the most common trigger criteria from
GBM, likely underestimating the sGRB rate by $20-30\%$ due to not
including the full range of energies and timescales in the robust GBM
triggering flight software.  

The BurstCube instrument is housed in a CubeSat bus based on the Dellingr platform,  a modular platform designed to be easily modifiable for a variety of 6U CubeSat architectures (see {\bf Fig. \ref{fig:spacecraft}}). It includes 3 axis-stabilized pointing control, Ultra-High Frequency (UHF) communication system, body mounted solar panels, and is designed for the Planetary Systems Corporation (PSC) 6U canister standard\footnote{Also compatible with the NanoRacks 6U dispenser}.  The {\it Dellingr} bus will be enhanced this year to include deployable solar panels and a star tracker.  The only additional  system needed for BurstCube is the Globalstar communications system for rapid GRB location distribution.  

\section{Performance}
{\bf The expected rate of coincident GW sources with sGRB detections with BurstCube is $\sim$2.2~$\textrm{yr}^{-1}$ for NS-NS progenitors,  and potentially significantly higher if sGRB progenitors include a BH.}   This estimate is drawn entirely from the observed sGRB rate, the GW detection range, and the intrinsic rate density of sGRBs in the nearby Universe and on the sensitivities of the detectors, but depends only weakly (quadratically) on the poorly-constrained beaming angle of sGRBs
\cite{2012ApJ...756..189F}.  BurstCube and other GRB-detecting observatories enhance the LIGO/Virgo detection rate by {\bf enabling the detection of weaker GW signals} due to their temporal and positional correlation with detected sGRBs. GW/sGRB rates are estimated by multiplying the search volume by an {\it
observed} sGRB rate of $10\pm5\ \textrm{Gpc}^{-3} \ \textrm{yr}^{-1}$
\cite{Swift_ShortGRBs,MNRA_ShortGRBs}. The search volume depends upon the makeup of the progenitor system, as the strength of the GW signal scales with the mass of the system, making binaries containing a BH detectable by LIGO/Virgo to larger distances.  A NS-BH binary increases the detection range compared to a NS-NS by a factor of $\sim 1.6$, and BH-BH by a factor of 2--5 \cite{2016arXiv160203842A}.  A factor of 1.5 is added due to the preferential enhancement of the GW signal along the jet axis, and another factor of 1.5 due to sGRB triggers seeding sub-threshold GW searches.  {\bf If sGRBs are also due to NS-BH mergers or BH-BH mergers, the sGRB/GW rate would increase by a factor of a few to hundreds.}  BurstCube is sensitive enough to detect the
$\gamma$ rays from any sGRB close enough for the GWs to be
detectable by LIGO/Virgo.

Because of the low expected rate of joint GW/sGRB detections, it is
crucial to detect sGRBs from the entire sky at all times.  Failure to
do so will lead to a very small sample of GW/sGRB coincident
observations, greatly weakening the scientific return of
LIGO/Virgo. 
{\bf BurstCube is designed as a low cost method to maximize sky coverage for detecting sGRBs to correlate with LIGO/Virgo.} 

{\bf BurstCube \emph{complements} existing and future GRB instruments} by
increasing the probability of coincident GW/sGRB detections, by increasing sky coverage.  GBM is the current best GRB
instrument for supporting the GW detectors, due to its very
wide field-of-view (FoV) and excellent detection sensitivity for sGRBs
with its 12 NaI and 2 BGO detectors \cite{2009ApJ...702..791M}.  However, any
similar observatory in a low Earth orbit can operate with at most $\sim60\%$
duty cycle for any point on the sky due to Earth blockage and South Atlantic
Anomaly (SAA) passage.  In comparison, {\it Swift}-BAT
\cite{2005SSRv..120..143B} views $\sim20\%$ of the sky at any one time, and the approved SVOM mission \cite{svom} has limited sky coverage and a soft energy band detecting $<10$ sGRBs yr$^{-1}$.  While the INTEGRAL SPI-ACS provides all-sky observations, it cannot localize or provide spectral information.  The Interplanetary Network (IPN) provides approximately all-sky coverage for bright sGRBs, but cannot probe fainter bursts, and has a data latency of up to several days. 
BurstCube will increase sky exposure by $\sim15\%$ on average depending on
orbit phasing.  The need for this is clear; even with the large FoVs of BAT and GBM the error contour of GW150914 was not fully covered at trigger  \cite{2016arXiv160208492A,2016arXiv160203920C,2016arXiv160203868E}.  
Even if {\it Swift} and {\it Fermi} continue to be operational in 2020, exposure improves with BurstCube as it provides a cost effective and timely method to
support and enhance LIGO/Virgo science as they reach design sensitivity.

\begin{figure}[b]
\begin{minipage}[c]{0.4\textwidth}
\includegraphics[width=\textwidth,trim=120 150 100 200,clip=true]{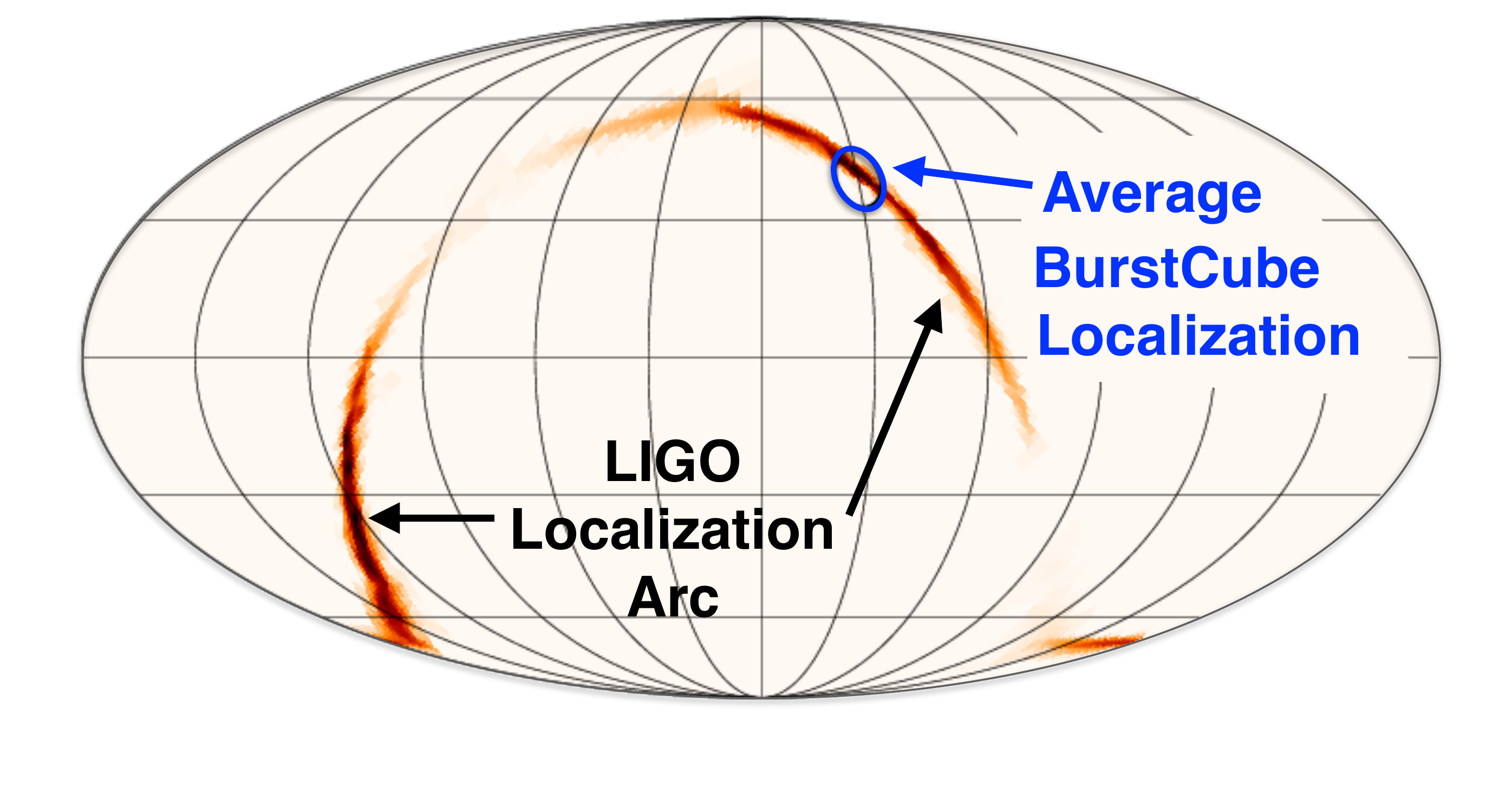}
\end{minipage}\hfill
\begin{minipage}[c]{0.6\textwidth}
  \caption{\small In some regions of the sky, the conjunction
    of BurstCube and LIGO/Virgo localizations will
    reduce the follow-up search area, especially in cases when the GW event is only seen by 2 detectors, as demonstrated by overlaying a simulated BurstCube localization over the LIGO localization map of GW151226 \cite{gw151226}.  The typical BurstCube localization (blue circle) is $\sim 7^{\circ}$ radius for the 2/3 of the unocculted sky with 3 or more detector exposures overlapping.
	\label{fig:localizations}}
\end{minipage}
\vspace{-4mm}
\end{figure}

While both LIGO/Virgo and BurstCube localizations on their own are
10's to 100's of deg$^2$, the intersection of GW and BurstCube
localizations ({\bf Fig. \ref{fig:localizations}}) for coincident
detections will reduce the area to be more manageable
for tiling. The reduction of these error regions will be especially important for events seen by only two GW detectors, sub-threshold GW signals, and those beyond the Virgo horizon. 

{\bf BurstCube will complement other observatories} by:
\begin{itemize}
\item providing higher energy
$\gamma$-ray coverage than {\it Swift}-BAT (15-150 keV) that can be used
in tandem to understand prompt emission spectra and estimate
energetics;
\item providing rapid ($\lesssim$5 minutes) localizations distributed via the Gamma-ray Coordinates Network (GCN) for afterglow searches by wide-FoV ground-based optical and space-based X-ray telescopes like ZTF \cite{2014htu..conf...27B}, MASTER \cite{2004AN....325..580L} and TAO-ISS (proposed) \cite{lobster};
\item adding to the IPN producing time-of-arrival
  localizations over long baselines.
\end{itemize}

BurstCube will also be sensitive to soft $\gamma$-ray phenomena including soft $\gamma$-ray repeaters, solar flares, and other transients.

As a CubeSat, the BurstCube detector sizes and quantity are limited
compared to GBM. GBM has a significantly wider energy range (8 keV - 30 MeV) due to the inclusion of two BGO detectors, each of which is more massive than a 6U
CubeSat.  However, several \textit{improvements} compared to GBM ameliorate
the size and number limitation. The use of CsI instead of NaI
provides better efficiency and more sensitivity at higher energies per unit
detector volume.  The smaller spacecraft will cause less scattering of
$\gamma$ rays, and that scattering will be easier to model, reducing
systematics in localizations.  Zenith (or Sun) pointing instead of the $50^\circ$ rocking of {\it Fermi} will simplify the modeling of $\gamma$ rays scattered by the Earth.
The BurstCube detectors will have passive
shielding to further reduce the effects of scattered $\gamma$ rays.  A faster
on board computer will allow a more sophisticated triggering algorithm optimized for sGRBs by using pre- and post-source background intervals.

\begin{figure}
\centering
\includegraphics[width=0.4\textwidth]{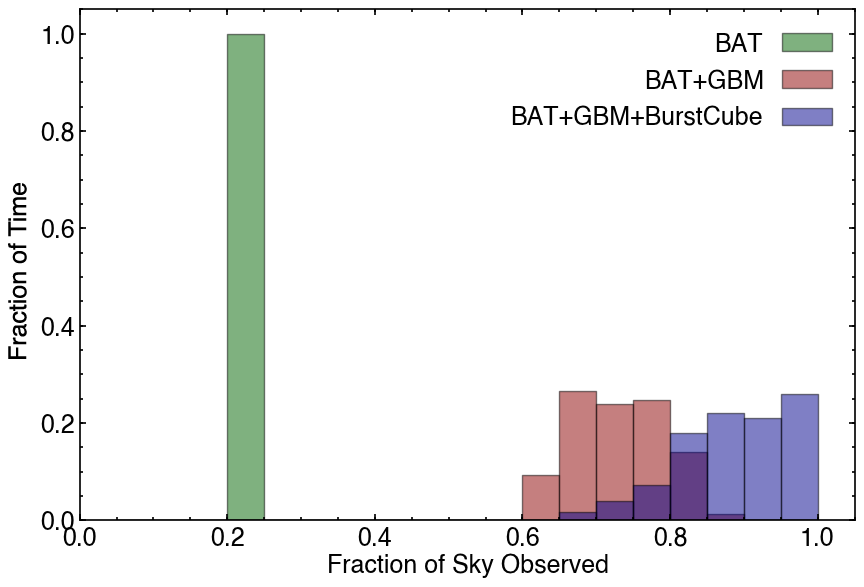}
\includegraphics[width=0.4\textwidth]{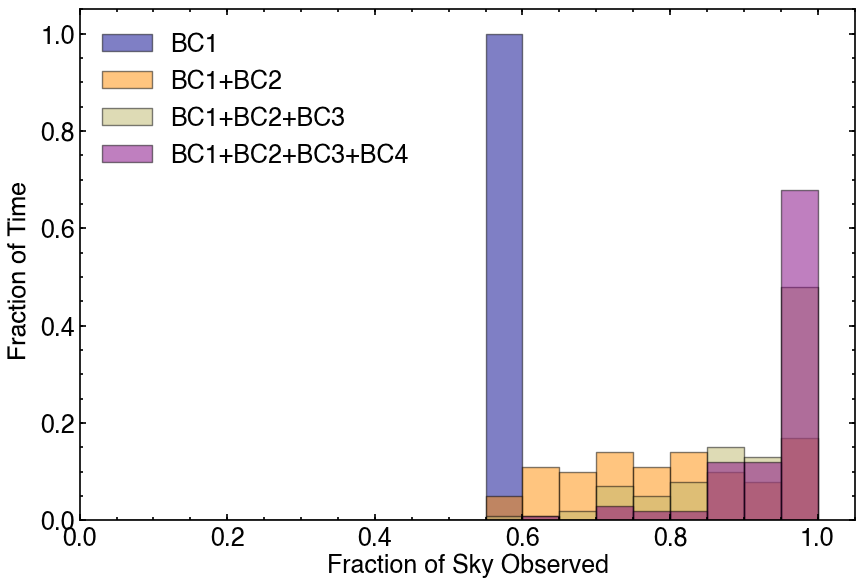}
\caption{\small Adding BurstCube to the existing network of GRB instruments {\it (left)} will increase sky coverage to improve the probability of detecting and characterizing coincident emission from a sGRB and GW trigger.  A set of 3 or more BurstCubes {\it (right)} would also provide all-sky coverage. \label{fig:skycov}}
\end{figure}

Accounting for the orbit phasing of BurstCube (assuming an ISS orbit), {\it Fermi}, and {\it Swift}, BurstCube increases the average sky coverage from  $\sim$70\% to $\sim$90\% ({\bf Fig. \ref{fig:skycov}}).
When LIGO/Virgo reach full sensitivity and joint GW/sGRB detections become a realistic prospect, it is vital that detectors sensitive to sGRBs with broad
sky coverage be operational.

\bibliographystyle{JHEP}
\bibliography{main}

\providecommand{\href}[2]{#2}\begingroup\raggedright\begin{thebibliography}{10}

\bibitem{gw150914}
B.~P. {Abbott} et~al., \emph{{Observation of Gravitational Waves from a Binary
  Black Hole Merger}},
  \href{http://dx.doi.org/10.1103/PhysRevLett.116.061102}{\emph{Physical Review
  Letters} {\bfseries 116} (Feb., 2016) 061102},
  [\href{https://arxiv.org/abs/1602.03837}{{\ttfamily 1602.03837}}].

\bibitem{gw151226}
B.~P. {Abbott} et~al., \emph{{GW151226: Observation of Gravitational Waves from
  a 22-Solar-Mass Binary Black Hole Coalescence}},
  \href{http://dx.doi.org/10.1103/PhysRevLett.116.241103}{\emph{Physical Review
  Letters} {\bfseries 116} (June, 2016) 241103},
  [\href{https://arxiv.org/abs/1606.04855}{{\ttfamily 1606.04855}}].

\bibitem{PhysRevLett.118.221101}
{\scshape LIGO Scientific and Virgo Collaboration} collaboration, B.~P. Abbott,
  R.~Abbott, T.~D. Abbott, F.~Acernese, K.~Ackley, C.~Adams et~al.,
  \emph{Gw170104: Observation of a 50-solar-mass binary black hole coalescence
  at redshift 0.2},
  \href{http://dx.doi.org/10.1103/PhysRevLett.118.221101}{\emph{Phys. Rev.
  Lett.} {\bfseries 118} (Jun, 2017) 221101}.

\bibitem{2016arXiv160203920C}
V.~{Connaughton} et~al., \emph{{Fermi GBM Observations of LIGO Gravitational
  Wave event GW150914}}, {\emph{ArXiv:1602.03920} (Feb., 2016) },
  [\href{https://arxiv.org/abs/1602.03920}{{\ttfamily 1602.03920}}].

\bibitem{2016arXiv160204542Z}
B.~{Zhang}, \emph{{Possible Short Gamma-Ray Bursts Associated with Black Hole -
  Black Hole Mergers}}, {\emph{ArXiv:1602.04542} (Feb., 2016) },
  [\href{https://arxiv.org/abs/1602.04542}{{\ttfamily 1602.04542}}].

\bibitem{2016arXiv160204460L}
X.~{Li}, F.-W. {Zhang}, Q.~{Yuan}, Z.-P. {Jin}, Y.-Z. {Fan}, S.-M. {Liu}
  et~al., \emph{{Implication of the association between GBM transient 150914
  and LIGO Gravitational Wave event GW150914}}, {\emph{ArXiv:1602.04460} (Feb.,
  2016) }, [\href{https://arxiv.org/abs/1602.04460}{{\ttfamily 1602.04460}}].

\bibitem{2016arXiv160204735L}
A.~{Loeb}, \emph{{Electromagnetic Counterparts to Black Hole Mergers Detected
  by LIGO}}, {\emph{ArXiv:1602.04735} (Feb., 2016) },
  [\href{https://arxiv.org/abs/1602.04735}{{\ttfamily 1602.04735}}].

\bibitem{2016arXiv160205050Y}
R.~{Yamazaki}, K.~{Asano} and Y.~{Ohira}, \emph{{Electromagnetic Afterglows
  Associated with Gamma-Ray Emission Coincident with Binary Black Hole Merger
  Event GW150914}}, {\emph{ArXiv:1602.05050} (Feb., 2016) },
  [\href{https://arxiv.org/abs/1602.05050}{{\ttfamily 1602.05050}}].

\bibitem{2016arXiv160205140P}
R.~{Perna}, D.~{Lazzati} and B.~{Giacomazzo}, \emph{{Short Gamma-Ray Bursts
  from the Merger of Two Black Holes}}, {\emph{ArXiv:1602.05140} (Feb., 2016)
  }, [\href{https://arxiv.org/abs/1602.05140}{{\ttfamily 1602.05140}}].

\bibitem{2013arXiv1304.0670L}
B.~P. {Abbott} et~al., \emph{{Prospects for Observing and Localizing
  Gravitational-Wave Transients with Advanced LIGO and Advanced Virgo}},
  \href{http://dx.doi.org/10.1007/lrr-2016-1}{\emph{Living Reviews in
  Relativity} {\bfseries 19} (Dec., 2016) 1},
  [\href{https://arxiv.org/abs/1304.0670}{{\ttfamily 1304.0670}}].

\bibitem{2013PhRvL.111g1101D}
W.~{Del Pozzo} et~al., \emph{{Demonstrating the Feasibility of Probing the
  Neutron-Star Equation of State with Second-Generation Gravitational-Wave
  Detectors}},
  \href{http://dx.doi.org/10.1103/PhysRevLett.111.071101}{\emph{Physical Review
  Letters} {\bfseries 111} (Aug., 2013) 071101},
  [\href{https://arxiv.org/abs/1307.8338}{{\ttfamily 1307.8338}}].

\bibitem{2010MNRAS.406.2650M}
B.~D. {Metzger} et~al., \emph{{Electromagnetic counterparts of compact object
  mergers powered by the radioactive decay of r-process nuclei}},
  \href{http://dx.doi.org/10.1111/j.1365-2966.2010.16864.x}{\emph{\mnras}
  {\bfseries 406} (Aug., 2010) 2650--2662},
  [\href{https://arxiv.org/abs/1001.5029}{{\ttfamily 1001.5029}}].

\bibitem{2009ApJ...702..791M}
C.~{Meegan}
  et~al.\href{http://dx.doi.org/10.1088/0004-637X/702/1/791}{\emph{\apj}
  {\bfseries 702} (Sept., 2009) 791--804},
  [\href{https://arxiv.org/abs/0908.0450}{{\ttfamily 0908.0450}}].

\bibitem{Megalib}
A.~{Zoglauer}, R.~{Andritschke} and F.~{Schopper}, \emph{{MEGAlib The Medium
  Energy Gamma-ray Astronomy Library}}, .

\bibitem{batse}
G.~J. {Fishman} et~al., \emph{{Overview of Observations from BATSE on the
  Compton Observatory}}, {\emph{\aaps} {\bfseries 97} (Jan., 1993) 17}.

\bibitem{2014ApJS..211...13V}
A.~{von Kienlin} et~al., \emph{{The Second Fermi GBM Gamma-Ray Burst Catalog:
  The First Four Years}},
  \href{http://dx.doi.org/10.1088/0067-0049/211/1/13}{\emph{\apjs} {\bfseries
  211} (Mar., 2014) 13}, [\href{https://arxiv.org/abs/1401.5080}{{\ttfamily
  1401.5080}}].

\bibitem{2012ApJ...756..189F}
W.~{Fong} et~al., \emph{{A Jet Break in the X-Ray Light Curve of Short GRB
  111020A: Implications for Energetics and Rates}},
  \href{http://dx.doi.org/10.1088/0004-637X/756/2/189}{\emph{\apj} {\bfseries
  756} (Sept., 2012) 189}, [\href{https://arxiv.org/abs/1204.5475}{{\ttfamily
  1204.5475}}].

\bibitem{Swift_ShortGRBs}
D.~{Guetta} and T.~{Piran}, \emph{The batse-swift luminosity and redshift
  distributions of short-duration grbs},
  \href{http://dx.doi.org/10.1051/0004-6361:20054498}{\emph{\aap} {\bfseries
  453} (2006) 823--828}.

\bibitem{MNRA_ShortGRBs}
D.~M. {Coward} et~al., \emph{{The Swift short gamma-ray burst rate density:
  implications for binary neutron star merger rates}},
  \href{http://dx.doi.org/10.1111/j.1365-2966.2012.21604.x}{\emph{\mnras}
  {\bfseries 425} (Oct., 2012) 2668--2673},
  [\href{https://arxiv.org/abs/1202.2179}{{\ttfamily 1202.2179}}].

\bibitem{2016arXiv160203842A}
B.~P. {Abbott} et~al., \emph{{The Rate of Binary Black Hole Mergers Inferred
  from Advanced LIGO Observations Surrounding GW150914}},
  {\emph{ArXiv:1602.03842} (Feb., 2016) },
  [\href{https://arxiv.org/abs/1602.03842}{{\ttfamily 1602.03842}}].

\bibitem{2005SSRv..120..143B}
S.~D. {Barthelmy} et~al., \emph{{The Burst Alert Telescope (BAT) on the SWIFT
  Midex Mission}},
  \href{http://dx.doi.org/10.1007/s11214-005-5096-3}{\emph{Space Science
  Reviews} {\bfseries 120} (Oct., 2005) 143--164},
  [\href{https://arxiv.org/abs/astro-ph/0507410}{{\ttfamily
  astro-ph/0507410}}].

\bibitem{svom}
D.~{G{\"o}tz} and {SVOM Collaboration}, \emph{{SVOM: a new mission for
  Gamma-Ray Bursts studies }}, {\emph{Memorie della Societa Astronomica
  Italiana Supplementi} {\bfseries 21} (2012) 162}.

\bibitem{2016arXiv160208492A}
B.~P. {Abbott} et~al., \emph{{Localization and broadband follow-up of the
  gravitational-wave transient GW150914}}, {\emph{ArXiv e-prints} (Feb., 2016)
  }, [\href{https://arxiv.org/abs/1602.08492}{{\ttfamily 1602.08492}}].

\bibitem{2016arXiv160203868E}
P.~A. {Evans} et~al., \emph{{Swift follow-up of the Gravitational Wave source
  GW150914}}, {\emph{ArXiv e-prints} (Feb., 2016) },
  [\href{https://arxiv.org/abs/1602.03868}{{\ttfamily 1602.03868}}].

\bibitem{2014htu..conf...27B}
E.~{Bellm}, \emph{{The Zwicky Transient Facility}},  in \emph{The Third
  Hot-wiring the Transient Universe Workshop} (P.~R. {Wozniak}, M.~J. {Graham},
  A.~A. {Mahabal} and R.~{Seaman}, eds.), pp.~27--33, 2014,
  \href{https://arxiv.org/abs/1410.8185}{{\ttfamily 1410.8185}}.

\bibitem{2004AN....325..580L}
V.~M. {Lipunov} et~al., \emph{{MASTER: The Mobile Astronomical System of
  Telescope-Robots}},
  \href{http://dx.doi.org/10.1002/asna.200410284}{\emph{Astronomische
  Nachrichten} {\bfseries 325} (Oct., 2004) 580--582},
  [\href{https://arxiv.org/abs/astro-ph/0411757}{{\ttfamily
  astro-ph/0411757}}].

\bibitem{lobster}
J.~{Camp} et~al., \emph{{Using ISS telescopes for electromagnetic follow-up of
  gravitational wave detections of NS-NS and NS-BH mergers}},
  \href{http://dx.doi.org/10.1007/s10686-013-9343-4}{\emph{Experimental
  Astronomy} {\bfseries 36} (Dec., 2013) 505--522},
  [\href{https://arxiv.org/abs/1304.3705}{{\ttfamily 1304.3705}}].

\end{thebibliography}\endgroup
\end{document}